\pdfoutput=1
\documentclass[11pt]{article}

\usepackage[preprint]{acl}

\usepackage{times}
\usepackage{latexsym}

\usepackage[T1]{fontenc}
\usepackage[utf8]{inputenc}

\usepackage{microtype}

\usepackage{inconsolata}

\usepackage{graphicx}

\usepackage{amsmath, amssymb}

\usepackage{booktabs} % For professional table formatting

\title{ClickAgent: Enhancing UI Location Capabilities of Autonomous Agents}

\author{
    Jakub Hoscilowicz*, Bartosz Maj, Bartosz Kozakiewicz \\ 
    {\bf Oleksii Tymoshchuk} \and {\bf Artur Janicki*} \\
    Samsung R\&D Poland \\ *Warsaw University of Technology
    }

\begin{document}
\maketitle
\footnotetext[1]{Correspondence\hspace{3pt}to:\hspace{2pt}\texttt{j.hoscilowic@samsung.com}, \texttt{artur.janicki@pw.edu.pl}.}

\begin{abstract}

With the growing reliance on digital devices equipped with graphical user interfaces (GUIs), such as computers and smartphones, the need for effective automation tools has become increasingly important. While multimodal large language models (MLLMs) like GPT-4V excel in many areas, they struggle with GUI interactions, limiting their effectiveness in automating everyday tasks. In this paper, we introduce ClickAgent, a novel framework for building autonomous agents. In ClickAgent, the MLLM handles reasoning and action planning, while a separate UI location model (e.g., SeeClick) identifies the relevant UI elements on the screen. This approach addresses a key limitation of current-generation MLLMs: their difficulty in accurately locating UI elements. ClickAgent outperforms other prompt-based autonomous agents (CogAgent, AppAgent) on the AITW benchmark. Our evaluation was conducted on both an Android smartphone emulator and an actual Android smartphone, using the task success rate as the key metric for measuring agent performance. The code for ClickAgent is available at \href{https://github.com/Samsung/ClickAgent}{\color{blue}{github.com/Samsung/ClickAgent}}.

\end{abstract}

\section{Introduction}

\begin{figure*}[t]
  \centering
  \includegraphics[scale=0.44]{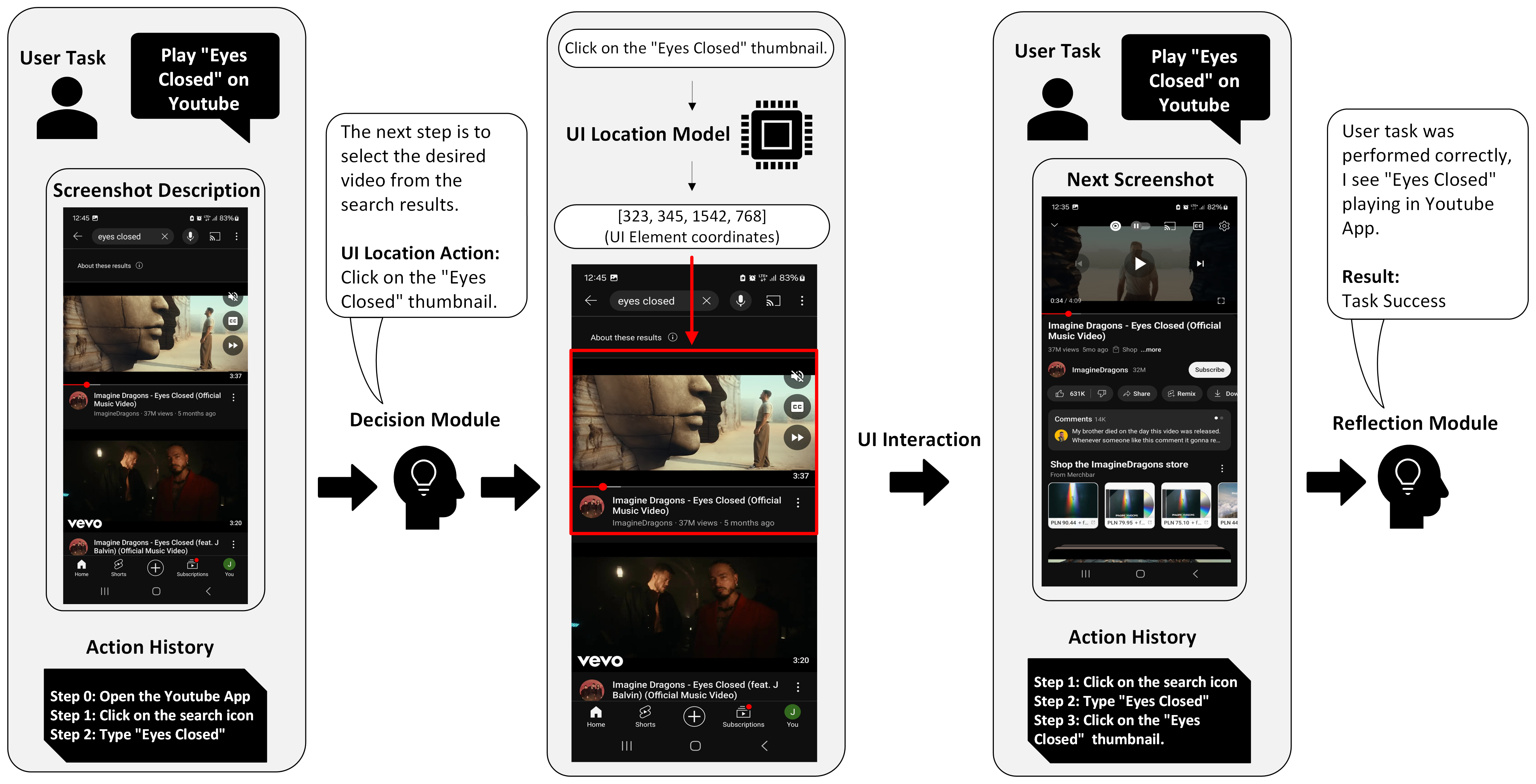}
  \caption{In ClickAgent, the MLLM is responsible for reasoning, reflection and action planning. In this example, the MLLM generates a UI command, and a specialized UI location model identifies the coordinates of the corresponding icon on the screen.}
  \label{fig:overview}
\end{figure*}

Autonomous agents capable of interacting with graphical user interfaces (GUIs) are becoming critical for automating tasks on digital devices such as smartphones and computers~\citep{kapoor2024omniact}. Researchers have begun developing agent-oriented large language models (LLMs)~\citep{chen2023fireact, zeng2023agenttuning}, but the potential of language-only agents is limited in real-world applications, where interaction with GUIs is often required. MLLMs and visual language models (VLMs) offer a promising solution to these limitations~\citep{you2023ferretrefergroundgranularity, rahman2024vzenefficientguiunderstanding, gur2024realworldwebagentplanninglong, baechler2024screenaivisionlanguagemodelui}. Unlike language-based agents that rely solely on textual data such as HTML~\citep{nakano2021webgpt} or OCR outputs \citep{rawles2023android}, MLLM-based agents directly interpret visual signals from GUIs. However, while current-generation MLLMs demonstrate reasonable abilities in screen understanding, reasoning, and action planning, they struggle to accurately locate specific UI elements on screens~\citep{liu2024visualwebbench}. Previous works~\citep{yang2023appagent, fan2024read, ma2024coco} attempt to bypass this issue, for example by using an XML file that details the interactive elements or by using a separate OCR model. Yet, such multi-module approaches are error-prone due to the inherent complexity of GUIs and the inconsistencies in XML/HTML files.

Our contribution lies in the development of ClickAgent, a hybrid autonomous agent that combines MLLM-driven reasoning with a specialized UI location model. Specifically, ClickAgent leverages the InternVL2.0 MLLM~\citep{chen2023internvl} for reasoning and the TinyClick UI location model for identifying the coordinates of relevant UI elements~\citep{pawlowski2024tinyclicksingleturnagentempowering}. This approach significantly advances performance, as demonstrated by our results on the commands from AITW benchmark~\citep{rawles2024androidinthewild}.

%asd
%asd2

\section{Method}

Although models like SeeClick \citep{cheng2024seeclick} and Auto-UI~\citep{zhan2023autoui} excel at identifying UI elements, they lack robust action planning, leading to low success rates in real-world smartphone tasks. To overcome these challenges, ClickAgent integrates InternVL2.0 for reasoning, while a dedicated UI location model identifies the exact coordinates of the target UI elements. The inputs to the UI location model are a screenshot and a natural language command corresponding to the desired UI element. ClickAgent's hybrid approach addresses the limitations of current MLLMs, which struggle to accurately locate UI elements \citep{liu2024visualwebbench}.

ClickAgent consists of three main components: Decision, UI Location and Reflection. In the Decision module, the MLLM is asked to analyze the current screenshot, review the action history, and determine the next step to complete the user’s task. The Decision module selects one of the predefined actions:

\begin{itemize}
    \item  $\mathtt{Click}$: The MLLM generates a natural language command for the UI location model (e.g., "click on the Gmail icon."). The command and screenshot are passed to the UI location model, which returns the coordinates of the relevant UI element.
    \item  $\mathtt{Type}$: The MLLM generates the text to be typed into the text field.
    \item  $\mathtt{Open App}$: If this action is chosen, an additional query is made to the MLLM to select an app from the list retrieved from the Android device.
    \item  $\mathtt{Swipe}$: The agent swipes in the specified direction (up, down, left, or right).
\end{itemize}

For example, as illustrated in Figure~\ref{fig:overview}, the MLLM chooses the Click action and issues a UI action command ("Click on the Eyes Closed Official Video") which, along with a screenshot, serves as input to the UI location model. The model then returns the bounding box coordinates of the relevant UI element. 

After the action is executed on Android, the next screenshot is captured. In the Reflection module, the MLLM is asked to analyze the content of screenshot and the entire action history. Reflection evaluates whether the user's task was successfully completed, returning either a "success" or "failure" status. If the task is marked as successful, the autonomous agent is stopped; otherwise, the agent proceeds with next decision and reflection cycle.

\begin{table*}[t]
    \centering
    %\small
    \begin{tabular}{lccc}
        \toprule
        \textbf{} & \textbf{AITW General} & \textbf{AITW WebShopping} & \textbf{Overall} \\
        \midrule
        AppAgent (GPT-4V)              & 15.6  & 13.5  & 14.0\\
        AutoUI                      & 12.5  & 18.8  & 17.2\\
        CogAgent                    & 25.0  & 42.6  & 38.3\\
        \midrule
        \textbf{ClickAgent (ours)}  & 72.5 & \textbf{75.8} & \textbf{73.5}  \\
        \hspace{5mm}\textit{+ Android cache removal (ours)} & \textbf{73.1} & 69.9 & 72.0 \\
        \bottomrule
    \end{tabular}
    \caption{Performance comparison of autonomous agents on the AITW General and AITW WebShopping benchmark subsets. Task success rates [\%] were calculated based on manual evaluations. The main ClickAgent results were obtained using an actual Android smartphone, while the results with cache removal were performed on an Android smartphone emulator to simulate a first-time user experience.}
    \label{tab:summary-results}
\end{table*}

\section{Evaluation Method}

We evaluated ClickAgent on both an Android smartphone emulator and a real Android smartphone, using the task success rate (in percentage) as the primary performance metric. This metric assesses whether the agent successfully executed the user's task, making it the most critical measure in autonomous agent evaluation. Unlike other metrics, such as step success rate or action accuracy \citep{zhan2023autoui}, the task success rate provides a clear and direct reflection of the agent's ability to accomplish user commands. Evaluation was performed manually due to imprecision of current automatic evaluation methods~\citep{pan2024autonomous}. We utilized  4 x NVIDIA A100 80GB GPU for running InternVL-2.0 and one NVIDIA A100 40GB GPU for the UI location models (SeeClick, TinyClick). Each experiment was repeated three times, showing no significant deviation in accuracy across runs. We report the average of these three runs.

\subsection{Test Environment}

The tests were conducted on both an Android smartphone and an Android emulator under two scenarios: with and without cache removal. In the cache removal scenario, the emulator's cache was cleared before each test case, ensuring that popups and first-time user interactions appeared for each website, simulating a first-time user experience. In the no-cache removal scenario, conducted on the real Android smartphone, the cache was retained to mimic a user who had previously visited the websites, thereby minimizing or eliminating popups and other initial distractions.

We conducted an evaluation of ClickAgent on a subset of the AITW dataset~\citep{rawles2023android}. In summary, our agent was evaluated on 154 unique webshop tasks and 432 general tasks. The AITW General consists of tasks related to interacting with everyday smartphone applications, while WebShopping focuses on tasks specific to e-commerce platforms.

\section{Main Results}

Table~\ref{tab:summary-results} presents the main results from the AITW benchmark. ClickAgent consistently outperforms other agents (AppAgent, Auto-UI, and CogAgent), achieving a significantly higher task success rate, regardless of whether the Android cache was cleared or not. As shown in Table~\ref{tab:ablation-results}, the accuracy of the UI location model plays a crucial role in determining the overall task success rate, making it a key factor in the ClickAgent's performance.

\subsection{UI Location Model Analysis}

Our primary insight is that TinyClick excels in OCR-related UI location. Therefore, we adjusted the prompt to encourage the MLLM to generate UI commands that incorporate textual information when possible. For instance, rather than producing commands like "Click on the first email," the MLLM is prompted to return more specific commands such as "Click on the email with the subject 'Meeting Agenda.'". This single prompt modification led to an improvement of around 10 percentage points in performance on the AITW.

\subsection{ClickAgent Fails Analysis}

On the AITW, the most common failures of ClickAgent were distributed across the following areas, with the percentages indicating the proportion of total errors attributed to each component:
\begin{itemize}
\item $\mathtt{Reflection\ Module}\ \mathtt{(47\%)}$: In some cases, the agent stops the action too late or too early, even though the task has not been completed.
\item $\mathtt{UI\ Location\ Model}\ \mathtt{(15\%)}$: Some UI elements are unique to specific applications, and certain web pages have outdated designs, making it challenging for the UI location model to accurately identify desired elements.
\item $\mathtt{Decision\ Module}\ \mathtt{(38\%)}$: Similar to the UI location model, most decision failures result from the MLLM's limited understanding of certain UIs and their functionalities.
\end{itemize}

Our experiments indicate that the performance of the Reflection and Decision is closely linked to the reasoning and general UI understanding capabilities of the MLLM. As more advanced MLLMs are developed, the performance of these modules should improve further.

\section{Ablation Study}

We conduct an ablation study to understand the impact of two main components (MLLM and UI location model) on the overall performance of ClickAgent. Table~\ref{tab:ablation-results} presents the evaluation of the UI location model's impact on ClickAgent's performance, comparing three different models. As expected, the MLLM (InternVL-2.0-76B) shows poor performance in the UI location task, resulting in ClickAgent failing all test cases. The most significant improvement comes from using the recently released TinyClick, which results in a substantially higher success rate than SeeClick.

\begin{table}[]
    \centering
    \footnotesize
    \setlength{\tabcolsep}{4pt} % Adjusts the space between columns
    \begin{tabular}{lcc}
        \toprule
        \textbf{} & \textbf{AITW General} & \textbf{AITW WebShopping}   \\
        \midrule
        InternVL2-76B              & 0  & 0  \\
        SeeClick-9.6B                      & 47.6  & 48.8  \\
        TinyClick-0.27B                    & 72.5  & 75.8  \\
        \bottomrule
    \end{tabular}
    \caption{Performance comparison of ClickAgent using different UI location models on the AITW General and AITW WebShopping (Task success rates [\%]).}
    \label{tab:ablation-results}
\end{table}

%\subsection{Impact of the used MLLM}

\begin{figure}[]
  \centering
  \includegraphics[scale=0.43]{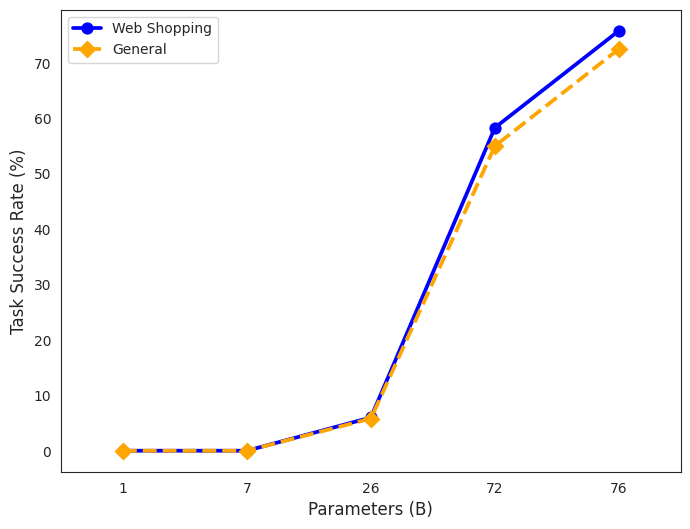}
      \caption{ClickAgent performance on the AITW General and AITW WebShopping using different MLLMs. In all cases, the TinyClick model is employed for UI location.}
  \label{fig:plot_models}
\end{figure}

Figure~\ref{fig:plot_models} illustrates the effect of MLLM general quality on ClickAgent's performance by evaluating four versions of InternVL-2.0 (1B, 7B, 26B, and 76B), alongside Qwen2.0-VL-72B~\citep{Qwen2VL}. The results show that the quality of the MLLM plays a critical role in the ClickAgent's performance. Larger models, such as InternVL-2.0-76B, result in significantly higher success rates compared to the smaller variants. Further improvements in the MLLM quality should continue to enhance the ClickAgent's performance (especially in terms of the accuracy of the Reflection module).

\section{Future Work}
%asd
Our failure analysis reveals that the constantly changing nature of UIs presents a significant challenge for autonomous agents. As UIs change, both MLLMs and UI location models require continuous retraining to maintain their ability to recognize and interact with new UI elements across apps and websites~\citep{chen2024guicourse, gao2024enhancing}. Another direction is the Retrieval-Augmented Generation (RAG) method introduced in AppAgent~\citep{yang2023appagent}. In this approach, apps and websites are first explored automatically, generating detailed documentation of their functionalities. During inference, RAG retrieves relevant sections from this documentation, acting as a real-time reference manual for the agent. Both continuous retraining and RAG are promising avenues for enhancing ClickAgent. Combined, they could enable the agent to better adapt to quickly changing UIs and improve its performance on less commonly encountered graphical interfaces.

\section{Conclusion}

In this paper, we introduced ClickAgent, a hybrid autonomous agent that combines MLLM-driven reasoning with a specialized UI location model. By addressing the limitations of previous approaches in identifying UI elements, ClickAgent demonstrates superior performance compared to baseline agents. The failures observed in the Reflection and Decision modules highlight the need for further advancements in MLLM capabilities, particularly in understanding UIs of less popular apps and websites.

%plus up to one page for limitations (required, see below) and optionally ethical considerations
\section*{Limitations}

The primary limitation of ClickAgent is its high task completion time, averaging around 60 sec. per task on the AITW benchmark. Such latency is a common issue among prompt-based autonomous agents~\citep{hong2023cogagent, yang2023appagent}, largely because the MLLM needs to generate detailed descriptions of screenshots. This prolonged inference time poses a challenge for latency-sensitive applications like voice assistants or other interactive AI systems. Figure~\ref{fig:plot_models} illustrates that while the smaller MLLMs have lower latency, their reasoning capabilities are too low for building a high-quality autonomous agent.

While alternative paradigms for building autonomous agents, such as Auto-UI, can reduce latency, they come with a significant trade-off in task success rates, as seen in Table~\ref{tab:summary-results}. These non-MLLM-based approaches prioritize speed but fail to match the multi-turn reasoning capabilities of prompt-based agents.

Reduction of MLLM inference time can be achieved through advancements in model efficiency, caching mechanisms, or specialized hardware. For instance, the NVIDIA H100 offers up to a 3x speed improvement over the A100, increasing throughput from 20 to 60 tokens per second. FPGA solutions, such as DFX, can push performance even further to 76 tokens per second~\citep{li2024large}. Another approach involves using real-time screen streaming (via MLLMs with video modalities) to take action shortly after screen changes (e.g., upon website load completion)~\citep{durante2024interactive}. Additionally, a promising direction is to hide the MLLM's reasoning and planning (e.g., screen description) using techniques like Implicit Chain-of-Thought~\citep{deng2024explicit, pfau2024let}.

\section*{Ethics}

Autonomous agents can be misused for malicious purposes, such as misinformation bots or automated surveillance. Implementing regulatory safeguards and raising societal awareness are essential for mitigating negative impacts. Furthermore, it is crucial to strengthen the robustness of autonomous agents against adversarial attacks from malicious actors~\citep{yang2024security}, so that benevolent agents are not manipulated into performing harmful actions.

\bibliography{_main}

\end{document}